# Effective Photon-Photon Interactions in Largely Detuned Optomechanics

Hao-Kun Li[†], Xue-Xin Ren, Yong-Chun Liu, and Yun-Feng Xiao[∗]

State Key Laboratory for Mesoscopic Physics and School of Physics, Peking University, Beijing 100871, P. R. China

We propose to realize effective beam-splitter-like and two-mode-squeezing photon-photon interactions in a strong coupling optomechanical interface by exploiting detuned driving lasers. In this interface, the transitions between the optical system and the mechanical oscillator are suppressed by the large energy offsets, therefore protecting the photon-photon interactions from mechanical dissipations. Moreover, the destructive quantum interference between the eigenmodes of the interface is capable of further reducing the effects of initial mechanical thermal occupations. The interface can serve as a universal block for photon state engineering and hybrid quantum networks in high-temperature thermal bath and without the requirement of cooling the mechanical oscillator to the ground state.

[†] Email address: hkli@pku.edu.cn

[∗] Email address: yfxiao@pku.edu.cn; URL: www.phy.pku.edu.cn/~yfxiao/index.html

Coherent photon-photon interactions are essentially desired in both studies of exotic quantum optical phenomena and various applications, ranging from optical quantum computation [1] and simulation [2] to quantum information processing [3]. Such interactions have been intensely studied in a variety of quantum systems, including atomic ensembles [4], cavity quantum electrodynamics [5], and surface plasmon polaritons [6]. Compared to them, benefiting from the capability of mechanical oscillators to couple to diverse optical fields, optomechanical systems open up great avenues in bridging the interactions between photons with vastly different wavelengths [7-17]. Experimentally, strong optomechanical couplings that exceed the cavity dampings have been demonstrated in both optical and microwave cavities [18-20], showing the great potential of optomechanical systems as media for photon state manipulations. Meanwhile, mechanical oscillators inherently suffer from thermal noise, which has been the main obstacle of generating and manipulating photon states in optomechanical interfaces. To overcome mechanical thermal dissipations in optomechanics, a common approach is to cool down the mechanical oscillator to its motional ground state [21-25]. However, the cooling processes themselves do not prevent thermal heating decoherence in quantum operations. A recently proposed approach is through the excitation of optomechanical dark mode which decouples from the mechanical oscillator [26-28]. By using the dark mode, photon state transfer can be performed with high fidelity against mechanical thermal dissipations.

In this paper, we propose to realize effective beam-splitter-like and two-mode-squeezing photon-photon interactions by employing detuned driving lasers, which provide a more universal platform for various quantum operations. During the optomechanical interactions, the transitions between the optical system and the mechanical oscillator are suppressed by the large energy offsets. Consequently, the interactions between photons are robust against mechanical dissipations. Moreover, the destructive quantum interference between the eigenmodes of the optomechanical system brings a significant advantage that photon state manipulations are insensitive to the initial mechanical states. This makes photon operations possible under large initial mechanical thermal occupations. For instance, we show that high fidelity photon state transfer and strong photon-photon entanglement can be realized in high-temperature thermal bath and without cooling the mechanical oscillator to the ground state.

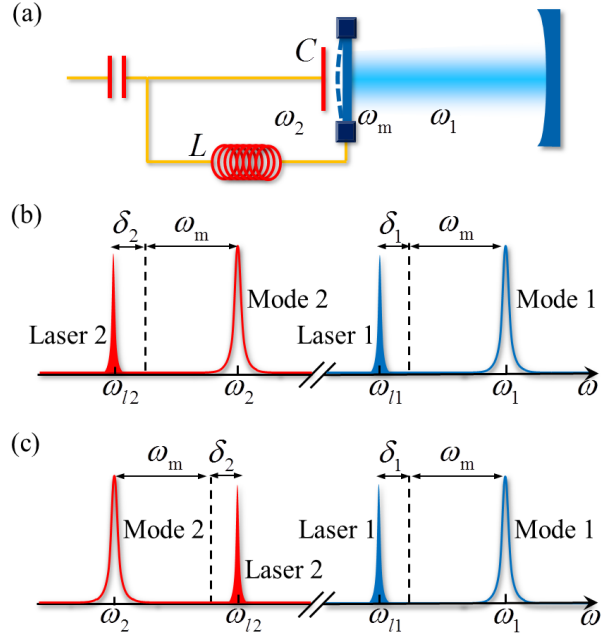

FIG. 1 (color online) (a) Schematic of the hybrid optomechanical interface. (b), (c) Two lasers drive the respective optomechanical couplings with detunings $-\Delta_i = \omega_m + \delta_i$ (b) and $-\Delta_1 = \omega_m + \delta_1$, $\Delta_2 = \omega_m + \delta_2$ (c).

The schematic of the hybrid optomechanical interface is shown in Fig. 1(a), where a mechanical mode is simultaneously coupled to an optical cavity mode and a microwave cavity mode via dispersive couplings. The two cavity modes are driven by strong lasers with frequencies $\omega_{l1}$ and $\omega_{l2}$, respectively. In the interaction picture with respect to the two driving lasers and in the displaced frame with respect to the average classical cavity fields, the Hamiltonian of the system under the standard linearization formalism can be written as ($\hbar = 1$)

$$H = \omega_m b^\dagger b - \sum_{i=1,2}[\Delta_i a_i^\dagger a_i - G_i(a_i^\dagger + a_i)(b^\dagger + b)]. \qquad (1)$$

Here $\omega_m$ and $b$ are the resonance frequency and annihilation operator of the mechanical mode; $a_i$ denotes the annihilation operator of cavity mode $i$ in the displaced frame; $G_i$ describes optically-driven coupling between the mechanical mode and cavity mode $i$; $\Delta_i = \omega_{li} - \omega_i$ represents the detuning of the drive applied to the cavity mode $i$ with $\omega_i$ being the optomechanically-shifted cavity resonance frequency. The damping rates of the mechanical mode and the cavity mode $i$ are denoted as $\gamma_m$ and $\kappa_i$. The environmental fluctuations can be represented by the cavity and mechanical input operators $a_{in}^{(i)}$ and $b_{in}$. In the high temperature limit, correlation functions of the input operators read $\langle a_{in}^{(i)}(t) a_{in}^{(i)\dagger}(t')\rangle = \delta(t-t')$ and $\langle b_{in}(t) b_{in}^{\dagger}(t')\rangle = (n_{th}+1)\delta(t-t')$ with $n_{th} = 1/(e^{\hbar\omega_m/k_B T}-1)$ being the environment thermal excitation phonon number at temperature $T$. In the following, we focus on the strong coupling regime, i.e., $G_i > \kappa_i, \gamma_m$. In addition, we assume that $\kappa_i \gg \gamma_m$, which is satisfied in typical optomechanical systems.

We first consider the case where both the cavity modes are driven near their red sidebands, with detunings $-\Delta_i = \omega_m + \delta_i$, as shown in Fig 1(b). In the limit $\omega_m \gg G_i, \delta_i$, the Hamiltonian of the system in the interaction picture can be simplified as $H_A = \sum_{i=1,2}[\delta_i a_i^\dagger a_i + G_i(a_i^\dagger b + a_i b^\dagger)]$ under the rotating-wave approximation, which describes the state exchanges between the mechanical mode and the two cavity modes. In this case, the quantum Langevin equation of the optomechanical interface can be written as $i\dot{\vec{v}}_A(t) = M_A \vec{v}_A(t) + i\sqrt{K}\vec{v}_{in}^A(t)$, with the operators $\vec{v}_A(t) = [a_1(t), b(t), a_2(t)]^T$, $\vec{v}_{in}^A(t) = [a_{in}^{(1)}(t), b_{in}(t), a_{in}^{(2)}(t)]^T$, and the matrices

$$M_A = \begin{pmatrix} \delta_1 - i\kappa_1/2 & G_1 & 0 \\ G_1 & -i\gamma_m/2 & G_2 \\ 0 & G_2 & \delta_2 - i\kappa_2/2 \end{pmatrix} \qquad (2)$$

and $K = \text{diag}[\kappa_1, \gamma_m, \kappa_2]$. For large detunings $\delta_i \gg G_i$, the transitions between the cavity system and the mechanical oscillator are suppressed by the large energy offsets [29,30]. By eliminating the mechanical mode, we obtain the effective beam-splitter-like interaction between the two cavity modes. The effective Hamiltonian reads

$$H_A^{eff} = \sum_{i=1,2}(\delta_i + \lambda_i)a_i^\dagger a_i + \lambda(a_1^\dagger a_2 + a_1 a_2^\dagger), \qquad (3)$$

where $\lambda = G_1 G_2 (\delta_1^{-1} + \delta_2^{-1})/2$ denotes the effective coupling strength between the two cavity modes, $\lambda_i = G_i^2/\delta_i$ describes the resonance AC-Stark shift of cavity mode $i$. The resonance condition of the beam-splitter-like interaction reads $\delta_1 + \lambda_1 = \delta_2 + \lambda_2$.

To quantitatively investigate the effects of the mechanical dissipations on the effective beam-splitter-like photon-photon interaction, we examine the full dynamics of the system, described by $\vec{v}_A(t) = U_A(t)\vec{v}_A(0) + \int_0^t U_A(t-t')\sqrt{K}\vec{v}_{in}^A(t')dt'$, where $U_A(t) = e^{-iM_A t}$. Here, we consider the two photon resonance condition $\delta_1 = \delta_2 = \delta$ and set the optomechanical couplings as $G_1 = G_2 = G$. For $\delta \gg G$ and $\lambda \gg |\kappa_1 - \kappa_2|$, the eigenmodes of $M_A$ are derived as $\varphi_1^A = \frac{1}{\sqrt{2}}[-1, 0, 1]^T$, $\varphi_2^A = \frac{1}{\sqrt{2}}[1, \frac{\sqrt{2}G}{\delta}, 1]^T$, and $\varphi_3^A = [\frac{G}{\delta}, 1, \frac{G}{\delta}]^T$, with corresponding eigenenergies $l_1^A = \delta - i\kappa/2$, $l_2^A = (\delta + 2\lambda) - i\kappa/2$, and $l_3^A = -2\lambda - i(\gamma_m + \frac{2G^2}{\delta^2}\kappa)/2$, respectively, where $\kappa = (\kappa_1 + \kappa_2)/2$. The eigenmodes $\varphi_1^A$ and $\varphi_2^A$ are mechanical dark mode doublets, which correspond to the eigenmodes of the effective beam-splitter-like Hamiltonian, while the eigenmode $\varphi_3^A$ is a mechanical bright mode. The matrix $U_A(t)$ can be expressed as

$$U_A(t) = \begin{pmatrix} e^{-itl^A}\cos(\lambda t) & s(t) & -ie^{-itl^A}\sin(\lambda t) \\ s(t) & e^{-itl_3^A} & s(t) \\ -ie^{-itl^A}\sin(\lambda t) & s(t) & e^{-itl^A}\cos(\lambda t) \end{pmatrix}, \quad (4)$$

with $s(t) = \frac{G}{\delta}(e^{-itl_2^A} - e^{-itl_3^A})$ and $l^A = \frac{1}{2}(l_1^A + l_2^A)$. It is noted that the cavity operators include a term $s(t)b(0)$ related to the initial mechanical state, and a term $\int_0^t s(t-t')\sqrt{\gamma_m}b_{in}(t')dt'$ with respect to the mechanical thermal heating. The effects of these terms on the covariance matrices (or the occupations) of the cavity modes are suppressed by a factor of $(G/\delta)^2$. In particular, when $(\delta + 4\lambda)t = 2q\pi$ ($q$ is integer), the destructive quantum interference between the eigenmodes $\varphi_2^A$ and $\varphi_3^A$ removes the initial mechanical state from the photon states, i.e., $s(t) \sim 0$. At these times, the initial mechanical thermal occupation plays a minor role in the cavity mode states. In Fig. 2, we plot the time evolution of the photon and phonon numbers $N_i(t) = \langle a_i^\dagger a_i \rangle_t$ and $N_b(t) = \langle b^\dagger b \rangle_t$ with initial occupations $N_1(0) = 1$, $N_2(0) = 0$, and $N_b(0) = 3$, under the average thermal phonon number of 0 and 150. As expected, the photon numbers exhibit Rabi oscillation with frequency $2\lambda$, under the envelope with respect to the averaged optical damping $\kappa$. The mechanical dissipations are greatly suppressed by both the large detuning and the destructive quantum interference, which can be verified by the fact that $N_1(t)$ with a slight oscillation is just above the value corresponding to $N_b(0) = 0$ and $n_{th} = 0$.

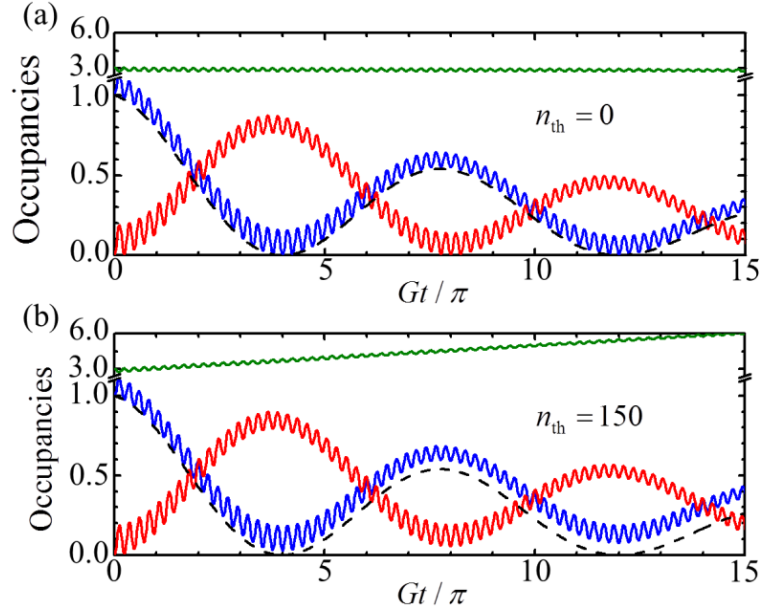

FIG. 2 (color online) Time evolution of $N_1$ (blue), $N_2$ (red), and $N_b$ (olive) for $n_{th} = 0$ (a) and $n_{th} = 150$ (b). Dashed black curves correspond to the evolution of $N_1$ at $N_b(0) = 0$ and $n_{th} = 0$. Parameters are $\omega_m = 50G$, $\kappa_1 = \kappa_2 = 0.025G$, $\gamma_m = 5\times 10^{-4}G$, and $\delta = 8G$.

The beam-splitter-like photon-photon interaction is at the heart of various quantum implementations, such as photon state transfer. Under the resonance condition $\delta_1 + \lambda_1 = \delta_2 + \lambda_2$, an intracavity photon state initially in cavity 1 can be transferred to cavity 2 by applying a $\pi/2$ pulse, i.e., $\lambda t_0 = \pi/2$. For simplicity, we assume that $G_1 = G_2 = G$, $\delta_1 = \delta_2 = \delta$, and have $\lambda > \kappa_i$. Consider the initial photon state to be the coherent state $|\alpha\rangle$, and the initial mechanical state to be the thermal state $\rho(n_m)$, with $n_m$ being the initial phonon number. Using the analytical expression of the cavity operator $a_2$, we derive the covariance matrix of the final state and obtain the fidelity of the photon state transfer $F = F_1 F_2$ (Ref. 31), where the intermediate fidelities are $F_1 = [1 + \frac{G^2}{\delta^2}(n_m | e^{-it_0 l_2^A} - e^{-it_0 l_3^A} |^2 + 2n_{th}\gamma_m t_0)]^{-1}$ and $F_2 = \exp[-(|\alpha|\kappa t_0/2)^2]$. It can be seen that effects of both the initial mechanical occupation and the thermal heating on the fidelity of the photon state transfer are reduced by the factor $(G/\delta)^2$. Moreover, when $(\delta + 4\lambda)t_0 = 2q\pi$, i.e., $\delta^2/4G^2 = q$, the final optical state is exempted from the initial mechanical state due to the quantum interference. Therefore, the fidelity remains high even for a large $n_m$, as shown in Fig. 3(a). In these cases, the fidelity $F_1$ is reduced to $[1 + \frac{G^2}{\delta^2}(n_m(\frac{\kappa-\gamma_m}{2}t_0)^2 + 2n_{th}\gamma_m t_0)]^{-1}$. In Fig. 3(b), we study the fidelity by varying the detuning $\delta$ in the condition of $\delta^2/4G^2 = q$. The results present a trade-off between thermal heating and cavity decays. Thus, for each temperature described by $n_{th}$, we can determine an optimal detuning and eventually derive the optimized fidelity of photon state transfer. The results are shown in Fig. 3(c). It can be seen that the present large detuning protocol has comparable high fidelity with the adiabatic state transfer [26], since the both are immune to the mechanical thermal heating unlike the double swap scheme [11]. Remarkably, the present optomechanical coupling system does not require precise modulation of the driving lasers, and provides a more general toolbox for different quantum operations.

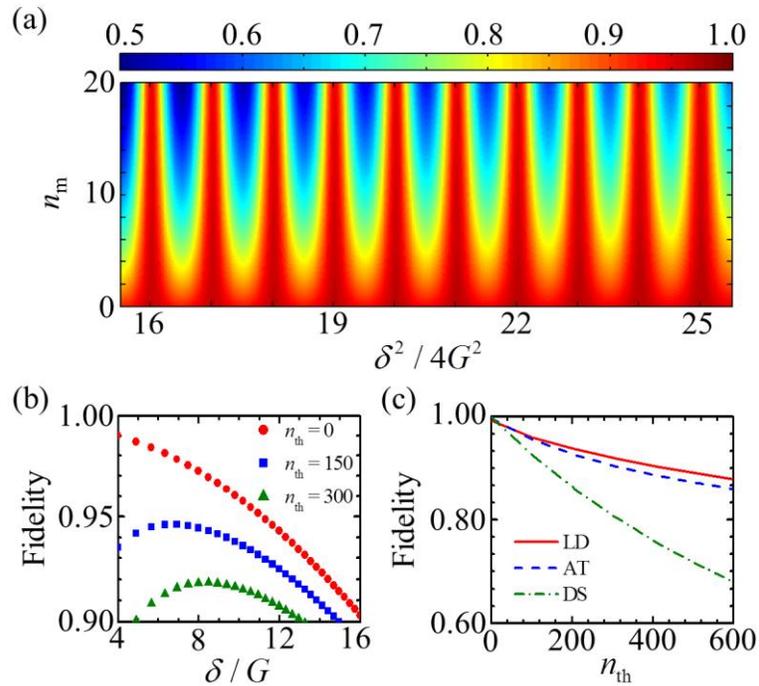

FIG. 3. (a) Intermediate fidelity $F_1$ as a function of detuning $\delta$ and initial phonon number $n_m$. Here $n_{th}$ = 200. (b) Fidelity $F$ versus detuning $\delta$ for $n_{th}$ = 0 (red circle), 150 (blue rectangle) and 300 (olive triangle). Here, $n_m$ = 20 and $\delta^2/4G^2 = q$. (c) Fidelity for photon state transfer using the large detuning (LD) protocol (solid red), the adiabatic dark state transfer (AT) (dashed blue), and the double swap (DS) (dashed dot olive) scheme. In (a)-(c), parameters are $\alpha = 1$, $\omega_m = 50G$, $\kappa_1 = \kappa_2 = 0.025G$, $\gamma_m = 5 \times 10^{-4}G$.

Next, we move to the situation where one of the cavity modes is driven near the red sideband $(-\Delta_1 = \omega_m + \delta_1)$, while the other is driven near the blue sideband $(\Delta_2 = \omega_m + \delta_2)$, as shown in Fig. 1(c). In the limit $\omega_m \gg G_i, \delta_i$, the Hamiltonian of the system in the interaction picture can be written as $H_B = (\delta_1 a_1^\dagger a_1 - \delta_2 a_2^\dagger a_2) + G_1(a_1^\dagger b + a_1 b^\dagger) + G_2(a_2 b + a_2^\dagger b^\dagger)$ under the rotating wave approximation. In such a case, the quantum Langevin equation of the system is given by $i\dot{\vec{v}}_B(t) = M_B \vec{v}_B(t) + i\sqrt{K}\vec{v}_{in}^B(t)$, with $\vec{v}_B(t) = [a_1(t), b(t), a_2^\dagger(t)]^T$, $\vec{v}_{in}^B(t) = [a_{in}^{(1)}(t), b_{in}(t), a_{in}^{(2)\dagger}(t)]^T$, and

$$M_B = \begin{pmatrix} \delta_1 - i\kappa_1/2 & G_1 & 0 \\ G_1 & -i\gamma_m/2 & G_2 \\ 0 & -G_2 & \delta_2 - i\kappa_2/2 \end{pmatrix}. \tag{5}$$

Similarly, for large detunings $\delta_i \gg G_i$, the direct interaction between the optical system and the mechanical mode is suppressed. Thus we obtain the effective two-mode-squeezing interaction between the two optical modes, described by the effective Hamiltonian

$$H_B^{eff} = \sum_{i=1,2} [\lambda_i - (-1)^i \delta_i] a_i^\dagger a_i + \lambda (a_1 a_2 + a_1^\dagger a_2^\dagger). \tag{6}$$

From Eq. (6), the parametric resonance condition is given by $\delta_1 + \lambda_1 = \delta_2 - \lambda_2$. In this case, a limit of $\lambda < \frac{1}{2}\sqrt{\kappa_1 \kappa_2}$ is required to make the system stable, thus hindering large effective coupling in quantum operations [32]. In the following, we consider the case where $\delta_1 = \delta_2 = \delta$, so that the photon-photon parametric resonance is avoided. Using the Routh-Hurwitz criterion [33], we find that the optomechanical system is stable when

$$G_1^2 \kappa_1 - G_2^2 \kappa_2 + \delta^2 \gamma_m > 0, \tag{7a}$$

$$(G_1^2 \kappa_1 - G_2^2 \kappa_2)(G_1^2 \kappa_2 - G_2^2 \kappa_1) + \delta^2 \kappa_1 \kappa_2 > 0. \tag{7b}$$

For $\delta = 0$, the stability condition is simplified into $G_1^2 / G_2^2 > \max(\kappa_1/\kappa_2, \kappa_2/\kappa_1)$, which requires $G_1 > G_2$ (Ref. 34). While for large detuning $\delta \gg G_i$, with proper optical and mechanical dampings, the system can be stable even when $G_1 \leq G_2$. This reveals that the large detuning is capable of enhancing the stability of the optomechanical system. In what follows, the optomechanical system is in the stable parameter regime described by Eqs. 7(a) and 7(b).

The effects of the mechanical dissipations on the effective two-mode-squeezing photon-photon interaction have also been studied quantitatively using the Langevin equation. In the condition $\delta \gg G_i > \kappa_1 = \kappa_2$, the eigenmodes of $M_B$ are expressed as $\varphi_1^B = [-\frac{G_2}{G_0}, 0, \frac{G_1}{G_0}]^T$, $\varphi_2^B = [\frac{G_1}{G_0}, \frac{\Delta\lambda}{G_0}, -\frac{G_2}{G_0}]^T$, and $\varphi_3^B = [\frac{G_1}{\delta}, 1, -\frac{G_2}{\delta}]^T$, with corresponding eigenenergies $l_1^B = \delta - i\kappa/2$, $l_2^B = \delta + \Delta\lambda - i\kappa/2$, and $l_3^B = -\Delta\lambda - i(\gamma_m + \frac{\Delta\lambda}{\delta}\kappa)/2$, respectively, where $G_0 = (G_1^2 + G_2^2)^{1/2}$ and $\Delta\lambda = \lambda_1 - \lambda_2$. The mechanical dark modes $\varphi_1^B$ and $\varphi_2^B$ correspond to the eigenmodes of the effective two-mode-squeezing Hamiltonian. In analogy to the case of the beam-splitter-like interaction, both the effects of the initial mechanical occupation and the mechanical thermal heating on the covariance matrices (or the occupations) of the cavity modes are suppressed by $(G_i/\delta)^2$. In parallel, the destructive quantum interference between the eigenmodes $\varphi_2^B$ and $\varphi_3^B$ cancels the initial mechanical state from the cavity states at the time satisfying $(\delta + 2\Delta\lambda)t = 2q\pi$.

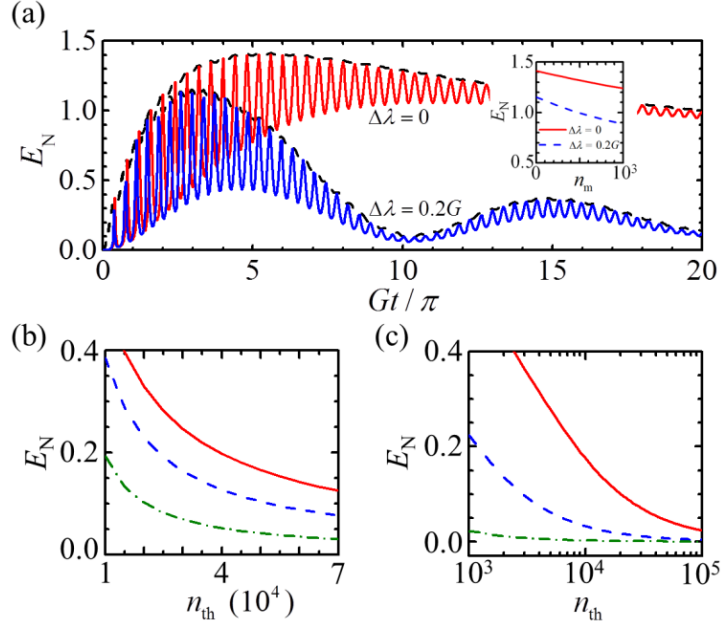

FIG. 4. (a) Time evolution of $E_N$ for $G_1 = G_2 = [\sinh(1)\cosh(1)]^{1/2}G$ (up red) and $G_1 = \cosh(1)G$, $G_2 = \sinh(1)G$ (bottom blue). Parameters are $\delta = 5G$, $n_m = 50$, and $n_{th} = 10^3$. Dashed black curves correspond to the $E_N$ with $n_m = 0$. Inset shows the maximum $E_N$ versus $n_m$. (b), (c) Maximum $E_N$ with $n_m = 10^3$ (b) and steady state $E_N$ (c) for $G_1 = G_2 = [\sinh(1)\cosh(1)]^{1/2}G$, $\delta = 15G$ (solid red), $G_1 = \cosh(1)G$, $G_2 = \sinh(1)G$, $\delta = 15G$ (dashed blue), and $G_1 = \cosh(1)G$, $G_2 = \sinh(1)G$, $\delta = 0$ (dashed dot olive). In (a), (b), and (c) $\omega_m = 50G$, $\kappa_1 = \kappa_2 = 0.025G$, $\gamma_m = 2\times 10^{-3}G$.

The two-mode-squeezing interaction is the key resource for generating continuous variable entanglement between the cavity fields, which is useful in many quantum information processes, such as quantum teleportation and cloning [17,35]. We quantify the photon-photon entanglement with the logarithmic negativity $E_N$, which can be obtained by deriving the covariance matrices of the optical system [36]. Assuming that initially the two cavity modes are in their vacuum states and the mechanical mode is in the thermal state $\rho(n_m)$, the dynamics of $E_N$ is plotted in Fig. 4(a). Because of the destructive quantum interference between the eigenmodes $\varphi_2^B$ and $\varphi_3^B$, $E_N$ reaches peak values of at the time with $(\delta + 2\Delta\lambda)t = 2q\pi$. At these times, the effects of the initial phonon occupation are significantly reduced, so that the instantaneous maximum $E_N$ for a large $n_m$ approaches the value at $n_m = 0$. The dependence of the maximum $E_N$ on $n_m$ is studied in the inset of Fig. 4(a). It is found that the maximum $E_N$ at $\Delta\lambda = 0$ ($\Delta\lambda = 0.2G$) shows a slight decrease from 1.40 (1.15) to 1.25 (0.90) when the initial phonon occupation $n_m$ increases from 0 to $10^3$. In addition, for $\Delta\lambda = 0.2G$, the interference between the dark mode doublets $\varphi_1^B$ and $\varphi_2^B$ results in a beat during the evolution of $E_N$. This process retards the generation of the photon-photon entanglement. In contrast, at the critical point $\Delta\lambda = 0$, the two dark modes become indistinguishable thus the beat disappears and stronger entanglement can be obtained. To verify the suppression of thermal heating, in Fig. 4(b) and 4(c), we plot the maximum $E_N$ and the steady state $E_N$, respectively, with different thermal excitation number $n_{th}$. It is shown that the photon-photon entanglement is much more robust against high temperature thermal bath compared to the case of zero detuning $\delta = 0$, which is considered in recent studies of photon-photon entanglement generation [17,34].

In summary, we have proposed to realize effective beam-splitter-like and two-mode-squeezing photon-photon interactions in a hybrid strong coupling optomechanical interface by employing largely detuned driving lasers. The photon-photon interactions are immune to both initial mechanical occupations and mechanical thermal heating due to the large energy offsets and the destructive quantum interference. The optomechanical interface enables various quantum implementations such as photon state transfer and photon-photon entanglement generation. It provides promising arenas for quantum state engineering and hybrid quantum networks.

## ACKNOWLEDGMENTS

This work is supported by NSFC (Nos. 11222440, 11004003, and 11121091), the 973 program (No. 2013CB328704 and No. 2013CB921904), and RFDPH (No. 20120001110068).

**References**

1. J. L. O'Brien, *Science* **318**, 1567 (2007).
2. M. J. Hartmann, F. G. S. L. Brandão, and M. B. Plenio, *Nature Physics* **2**, 849 (2006).
3. D. Bouwmeester, A. Ekert, and A. Zeilinger, *The Physics of Quantum Information* (Springer, Berlin, 2000).
4. M. D. Lukin, *Rev. Mod. Phys.* **75**, 457 (2003).
5. K. M. Birnbaum, A. Boca, R. Miller, A. D. Boozer, T. E. Northup and H. J. Kimble, *Nature* **436**, 87 (2005).
6. D. E. Chang, A. S. Sørensen, E. A. Demler, and M. D. Lukin, *Nature Physics* 3, 807 (2007).
7. T. J. Kippenberg and K. J. Vahala, *Science* **321**, 1172 (2008).
8. F. Marquardt and S. M. Girvin, *Physics* **2**, 40 (2009).
9. S. Weis, R. Rivière, S. Deléglise, E. Gavartin, O. Arcizet, A. Schliesser, and T. J. Kippenberg, *Science* **330**, 1520 (2010).
10. A. H. Safavi-Naeini, T. P. M. Alegre, J. Chan, M. Eichenfield, M. Winger, Q. Lin, J. T. Hill, D. E. Chang, and O. Painter, *Nature* **472**, 69 (2011).
11. L. Tian and H. Wang, *Phys. Rev. A* **82**, 053806 (2010).
12. V. Fiore, Y. Yang, M. C. Kuzyk, R. Barbour, L. Tian, and H. Wang, *Phys. Rev. Lett.* **107**, 133601 (2011).
13. J. T. Hill, A. H. Safavi-Naeini, J. Chan, and O. Painter, *Nature Comm.* **3**, 1196 (2012).
14. T. A. Palomaki, J.W. Harlow, J. D. Teufel, R.W. Simmonds, and K.W. Lehnert, *Nature* **495**, 210 (2013).
15. M. Paternostro, D. Vitali, S. Gigan, M. S. Kim, C. Brukner, J. Eisert, and M. Aspelmeyer, *Phys. Rev. Lett.* **99**, 250401 (2007).
16. Sh. Barzanjeh, D. Vitali, P. Tombesi, and G. J. Milburn, *Phys. Rev. A* **84**, 042342 (2011).
17. Sh. Barzanjeh, M. Abdi, G. J. Milburn, P. Tombesi, and D. Vitali, *Phys. Rev. Lett.* **109**, 130503 (2012).

18. S. Gröblacher, K. Hammerer, M. R. Vanner, and M. Aspelmeyer, *Nature* **460**, 724 (2009).
19. J. D. Teufel, D. Li, M. S. Allman, K. Cicak, A. J. Sirois, J. D. Whittaker, and R. W. Simmonds, *Nature* **471**, 204 (2011).
20. E. Verhagen, S. Deléglise, S. Weis, A. Schliesser, and T. J. Kippenberg, *Nature* **482**, 63 (2012).
21. J. D. Teufel, T. Donner, D. Li, J. W. Harlow, M. S. Allman, K. Cicak, A. J. Sirois, J. D. Whittaker, K. W. Lehnert, and R. W. Simmonds, *Nature* **475**, 359 (2011).
22. J. Chan, T. P. M. Alegre, A. H. Safavi-Naeini, J. T. Hill, A. Krause, S. Gröblacher, M. Aspelmeyer, and O. Painter, *Nature* **478**, 89 (2011).
23. X. Wang, S. Vinjanampathy, F. W. Strauch, and K. Jacobs, *Phys. Rev. Lett.* **107**, 177204 (2011).
24. S. Machnes, J. Cerrillo, M. Aspelmeyer, W. Wieczorek, M. B. Plenio, and A. Retzker, *Phys. Rev. Lett.* 108, 153601 (2012).
25. Y.-C. Liu, Y.-F. Xiao, X. Luan, and C. W. Wong, *Phys. Rev. Lett.* **110**, 153606 (2013).
26. Y.-D. Wang and A. A. Clerk, *Phys. Rev. Lett.* **108**, 153603 (2012).
27. L. Tian, *Phys. Rev. Lett.* **108**, 153604 (2012).
28. C. Dong, V. Fiore, M. C. Kuzyk, and H. Wang, *Science* **338**, 1609 (2012).
29. J. I. Cirac, P. Zoller, H. J. Kimble, and H. Mabuchi, *Phys. Rev. Lett.* **78**, 3221 (1997).
30. S.-B. Zheng and G.-C.Guo, *Phys. Rev. Lett.* **85**, 2392 (2000).
31. A. Isar, *Eur. Phys. J. Special Topics* **160**, 225 (2008).
32. C. Genes, A. Mari, P. Tombesi, and D. Vitali, *Phys. Rev. A* **78**, 032316 (2008).
33. E. X. DeJesus and C. Kaufman, *Phys. Rev. A* **35**, 5288 (1987).
34. L. Tian, *arXiv*:1301.5376 (2013)
35. S. L Braunstein and P. van Loock, *Rev. Mod. Phys.* **77**, 513 (2005).
36. D. Vitali, S. Gigan, A. Ferreira, H. R. Böhm, P. Tombesi, A. Guerreiro, V. Vedral, A. Zeilinger, and M. Aspelmeyer, *Phys. Rev. Lett.* **98**, 030405 (2007).